\documentclass[aps,prl,twocolumn,superscriptaddress]{revtex4}\newcommand{\figscale}{.7} 

\usepackage{graphicx}
\usepackage{xcolor}
\usepackage{xspace}
\usepackage{dsfont}
\usepackage{amsfonts}
\usepackage{amsmath}
\usepackage{amssymb}
\usepackage{axodraw2}
\usepackage[colorlinks=true, linkcolor=blue, citecolor=blue, urlcolor=blue]{hyperref}
\usepackage{makerobust}
\MakeRobustCommand\scalebox
\def\bi{\begin{itemize}}
\def\ei{\end{itemize}}
\def\be{\begin{equation}}
\def\ee{\end{equation}}
\def\bea{\begin{eqnarray}}
\def\eea{\end{eqnarray}}
\newcommand{\tn}{\textnormal}

\def\eg{e.\,g.}
\def\ie{i.\,e.}
\def\wrt{w.\,r.\,t.}

\def\gsim{\mathrel{\rlap{\lower0.2em\hbox{$\sim$}}\raise0.2em\hbox{$>$}}}
\def\lsim{\mathrel{\rlap{\lower0.2em\hbox{$\sim$}}\raise0.2em\hbox{$<$}}}
\def\lg{\mathrel{\rlap{\lower0.25em\hbox{$>$}}\raise0.25em\hbox{$<$}}}

\def\xzero{\ensuremath{x_{\scalebox{0.6}{0}}}}
\def\xone{\ensuremath{x_{\scalebox{0.6}{1}}}}
\def\xtwo{\ensuremath{x_{\scalebox{0.6}{2}}}}
\def\xF{\ensuremath{x_{\scalebox{0.6}{F}}}} 

\def\qhat{\hat{q}}

\newcommand{\eq}[1]{(\ref{#1})}
\newcommand{\R}{{\rm R}}

\newcommand{\pt}{p_{_\perp}}
\newcommand{\pp}{\ensuremath{\text{pp}}\xspace}
\newcommand{\pA}{\ensuremath{\text{pA}}\xspace}
\newcommand{\dd}{{\rm d}}

\def\bm#1{\mbox{\boldmath$#1$}}

\begin{document}

\title{Depletion of atmospheric neutrino fluxes from parton energy loss} 

\author{Fran\c{c}ois Arleo}
\email{francois.arleo@cern.ch}
\affiliation{Laboratoire Leprince-Ringuet, \'Ecole polytechnique, Institut polytechnique de Paris, CNRS/IN2P3, Route de Saclay, 91128 Palaiseau, France}
\affiliation{SUBATECH UMR 6457 (IMT Atlantique, Universit\'e de Nantes, IN2P3/CNRS), 4 rue Alfred Kastler, 44307 Nantes, France}
\author{Greg Jackson}
\email{gsj6@uw.edu}
\affiliation{Institute for Nuclear Theory, Box 351550, University of Washington, Seattle, WA 98195-1550, United States}
\author{St\'ephane Peign\'e}
\email{peigne@subatech.in2p3.fr}
\affiliation{SUBATECH UMR 6457 (IMT Atlantique, Universit\'e de Nantes, IN2P3/CNRS), 4 rue Alfred Kastler, 44307 Nantes, France}

\begin{abstract}
The phenomenon of fully coherent energy loss (FCEL) in the collisions of protons on light ions affects the physics of cosmic ray air showers. 
As an illustration, we address two closely related observables: 
hadron production in forthcoming proton-oxygen collisions at the LHC, 
and the atmospheric neutrino fluxes induced by the semileptonic decays 
of hadrons produced in proton-air collisions. 
In both cases, a significant nuclear suppression due to FCEL is predicted. 
The conventional and prompt neutrino fluxes are suppressed by $\sim 10...25$\% 
in their relevant neutrino energy ranges. Previous estimates of atmospheric neutrino fluxes should be scaled down accordingly to account for FCEL. 
\end{abstract}

\maketitle

%%%%%%%%%%   Introduction  %%%%%%%%%%%%%%

When a cosmic ray (CR), usually a proton, impacts the Earth's atmosphere, it 
collides with a nucleus of average mass number $\left< A \right> \simeq 14.5$ and 
initiates a collimated shower of particles, amid which the incident energy is dispersed. 
During the formation of the air shower, the particles having 
semileptonic decay modes 
generate an {\em atmospheric} neutrino flux that constitutes an important background 
to the neutrinos of astrophysical origin being hunted by several 
experiments~\cite{Abbasi:2010qv,Aartsen:2013jdh,Aartsen:2016xlq,Adamson:2012gt,Adrian-Martinez:2013bqq,Albert:2021pwz,Richard:2015aua}. 
The inclusive flux of atmospheric neutrinos receives two 
contributions, named the `conventional' flux arising from decays of long-lived particles (mostly $\pi$ and $K$ mesons), 
and the `prompt' flux due to decays of short-lived particles (mostly $D$ mesons). 
At very high neutrino energy $E_{\nu}$, the prompt source overcomes the conventional one, 
the transition energy between the two regimes being 
in the range $E_{\nu} = 10^4...10^6$~GeV, depending on the CR's zenith angle~\cite{Gondolo:1995fq,Fedynitch:2018cbl}.

Reaching a better understanding of cosmic ray air showers is one of the reasons 
to plan a proton-oxygen (pO) run at the LHC~\cite{Brewer:2021kiv}, whose foreseen collision energy, 
$\sqrt{s_{_\tn{NN}}}(\tn{pO}) = 9.9$~TeV, 
translates into a CR proton energy $E_\tn{p} = 5.2\times 10^7$~GeV in the oxygen rest frame. 
The physics of air showers is related to particle production at {\it forward} rapidities. 
In this respect, fully coherent energy loss (FCEL) is 
an important effect~--~so far omitted in previous air shower studies. 
FCEL has been derived in various 
formalisms~\cite{Arleo:2010rb,Armesto:2012qa,Armesto:2013fca,Peigne:2014uha,Peigne:2014rka,Liou:2014rha,Munier:2016oih} 
and plays a crucial role in quarkonium~\cite{Arleo:2012hn,Arleo:2012rs,Arleo:2013zua}, 
light hadron~\cite{Arleo:2020eia,Arleo:2020hat} 
and heavy meson~\cite{Arleo:2021bpv} nuclear suppression in pA collisions. 
To illustrate this point, let us first discuss hadron 
($h$) nuclear suppression expected from FCEL, 
in pO collisions at $\sqrt{s}= 9.9$~TeV. 
%%%%%%%%%%%%%%%%%%%%%%%%%%%%%%%%%%%%
\begin{figure}[t]
  \includegraphics[scale=\figscale]{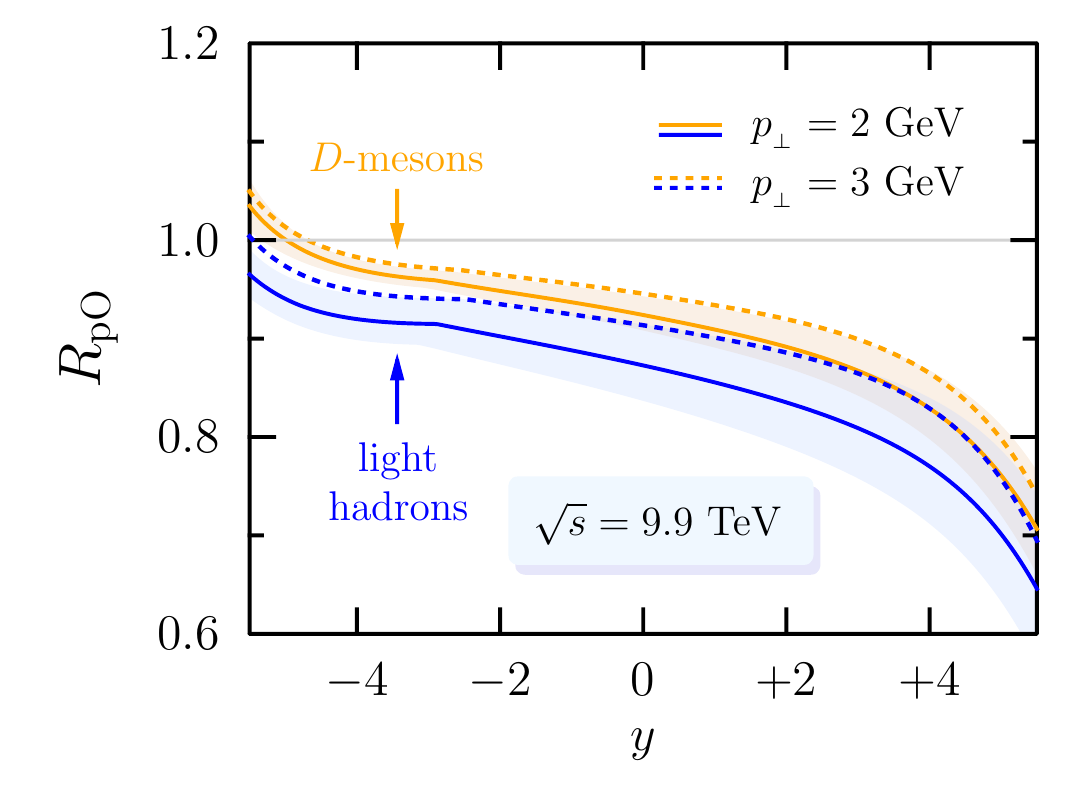}
  \vskip -3mm
  \caption{\label{fig:RpO} 
  Rapidity dependence of $D$ meson and light hadron nuclear suppression due solely to FCEL, 
  in pO collisions at $\sqrt{s} = 9.9$ TeV and for different values of $\pt$. 
  The uncertainty bands (shown only for $\pt=2$~GeV) are obtained 
  as in Refs.~\cite{Arleo:2020eia,Arleo:2020hat, Arleo:2021bpv} for pPb collisions. 
  }
\end{figure}
%%%%%%%%%%%%%%%%%%%%%%%%%%%%%%%%%%%%%
Before elaborating on the model, we display in Fig.~\ref{fig:RpO} the nuclear modification factor  
\be
R_{\pA}^{h}\big(\,y,\,\pt;\sqrt{s}\,\big)
  = 
  \frac{1}{A} \, {\frac{\dd\sigma_{\pA}^{h}}{\dd y \, \dd \pt} 
  \bigg/
  \frac{\dd\sigma_{\pp}^{h}}{\dd y \, \dd \pt}}  \, ,
\label{eq:R_pA}
\ee
obtained for $D$ mesons and light hadrons, as a function of the 
rapidity $y$ and for different values of the transverse momentum $\pt$. 
We observe that hadron suppression due to FCEL increases with $y$ (at fixed $\pt$), 
and becomes substantial at forward rapidities. As in the case of 
proton-lead (pPb) collisions~\cite{Arleo:2020eia,Arleo:2020hat,Arleo:2021bpv}, FCEL 
needs to be considered for hadron production in forthcoming pO collisions at the LHC, 
despite the smaller size of the oxygen nucleus~\footnote{%
  This can be understood from the mild dependence on $A$ in the 
  average FCEL, namely, $\Delta E_{_\tn{FCEL}} \propto A^{1/6}$~\cite{Arleo:2010rb}.
}.
It stands to reason, therefore, that FCEL should modify the structure of CR air showers. 
In this Letter we show that the implementation of FCEL 
in the {\it inclusive} neutrino flux calculation backs up the main point suggested by Fig.~\ref{fig:RpO}, 
namely, FCEL significantly suppresses both kinds of atmospheric neutrino fluxes. 

%%%%%%%%%%%      FCEL model details    %%%%%%%%%%%%%%%

Let us first recap the FCEL setup in pA collisions. 
The fully coherent induced gluon radiation associated with 
the underlying partonic subprocess has the effect 
of scaling down the hadron energy \wrt~the pp case, 
by a factor $z$ with a probability distribution ${\cal F}(z)$ called the {\em quenching weight}. 
At leading-order (LO) in $\alpha_s$, the hadron production 
cross section is given by $2 \to 2$ partonic processes
and the final partons are thus back-to-back in the transverse plane.
We will denote by $K_{\perp}$ and $\xi$ the transverse momentum and energy fraction of the parton that fragments 
into the hadron, 
and implement FCEL in the approximation where the final parton pair, of invariant mass 
$M \equiv M(K_\perp, \xi)$, behaves as a pointlike object with regard to the induced radiation. 

Within this `pointlike dijet approximation' (PDA)~\footnote{%
  The PDA can be shown to hold within the logarithmic accuracy 
  $\log{(\frac{\qhat L_\tn{A}}{x^2 M^2})} \gg 1$~\cite{Arleo:2020hat,Arleo:2021bpv}, 
  which turns out to be well satisfied in the present study.
}, 
the induced spectrum ${\dd{I}_{_\R}}/{\dd x}$ 
(with $x$ the fractional energy loss) for a parton pair in colour state $\R$ 
(of Casimir $C_\R$) reads~\cite{Arleo:2020eia,Arleo:2020hat,Arleo:2021bpv}
\be
x \frac{\dd I_{_\R}}{\dd x}  = \big( \, C_1 + C_\R - C_2 \, \big) \, \frac{\alpha_s}{\pi} \, 
\log{\frac{\qhat \, L_\tn{A} + x^2 M^2}{\qhat \, L_\tn{p} + x^2 M^2}} \, .
\label{dIR} 
\ee
Here $C_1$ and $C_2$ are the (Casimir) 
charges of the incoming partons from the proton and target nucleus, respectively,  
and $\sqrt{\qhat L_\tn{A}}$ is the transverse momentum `kick' 
imparted to the energetic incoming parton by the nuclear target, 
with $L_\tn{A}$ the distance travelled across the target and ${\hat q}$ the cold nuclear matter transport coefficient.

The FCEL quenching weight for a given partonic process and colour state $\R$ 
is conveniently expressed in terms of the energy rescaling factor $z \equiv 1/(1+x)$ as~\cite{Arleo:2020hat} 
\be
{\cal F}_{_\R}(z) =
- \, \frac{\partial}{\partial z} \, \exp
\bigg[
    \int_{0}^z \dd {z^{\prime}}  \, \frac{\dd I_{_\R}}{\dd {z^{\prime}}}
\bigg]  \, ,
\label{eq:F(z)} 
\ee
which represents a normalised probability distribution,
viz.~$\int_0^1 \dd z \, {\cal F}_{_\R}(z) = 1$. 
In the target nucleus rest frame, a rescaling of energy is equivalent to a rescaling of light-cone momentum $p^+ \equiv E + p^z$. 
Moreover, in the PDA the same energy rescaling applies to the incoming parton, the two partons of the final state,
and the produced hadron. 
Thus, FCEL can be formulated as a rescaling of 
the Feynman variable $\xF \equiv p_h^+/p_\tn{p}^+$ when comparing 
the hadron production cross section in pp and pA collisions, namely~\cite{Arleo:2012rs}, 
\bea
\frac1{A} \frac{\dd \sigma_{\tn{pA}}^h}{\dd \xF}
\big( \xF; E_\tn{p} \big) &=&  \int_{\xF}^1 \! \dd z \, {\cal F}(z) \,
\frac{\dd \sigma_\tn{pp}^h}{\dd \xF}
\bigg( \frac{\xF}{z} ; E_\tn{p} \bigg) \, .  \quad
\label{eq:pA}
\eea
Here ${\cal F}(z) = \sum_\R \rho_{_\R}(\xi) {\cal F}_{_\R}(z)$, accounting for the fact that each colour state $\R$ comes with 
a probability $\rho_{_\R}$ (that depends only on $\xi$), 
and we traded $\sqrt{s}$ for $E_\tn{p} \simeq s / (2 m_\tn{p})$, where $m_\tn{p}$ is the proton mass. 

Figure~\ref{fig:RpO} is obtained from Eq.~\eq{eq:R_pA} by using \eq{eq:pA} adapted to
$\dd \sigma / \dd y $ after changing variables from $\xF$ to $y$ 
and given a parametrization of the differential pp cross section (see~\cite{Arleo:2020hat,Arleo:2021bpv}). 
At the LHC, because of the largeness of the gluon distribution function, the dominant LO partonic process for $D$ meson 
(respectively,~light hadron) production is $gg \to c \bar{c}$ (respectively,~$gg \to gg$), 
with available colour states $\R = \{\bm 1, \bm 8\}$ (respectively,~$\R = \{\bm 1, \bm 8, \bm{27} \}$) 
of probability $\rho_{_\R}(\xi)$ derived previously~\cite{Arleo:2020hat,Arleo:2021bpv}. 
Then the parton pair invariant mass reads $M^2 = m_{\perp}^2/\bm(\xi(1-\xi)\bm)$ with $m_\perp^2 \equiv K_\perp^2 + m^2$, 
where $m=m_c\simeq1.3$~GeV~\cite{Yao:2006px} or $m=0\,$. 
The longitudinal momentum fractions $\xone$ and $\xtwo$ of the incoming gluons are 
given by $\xone = m_{\perp} e^{y}/(\xi \sqrt{s}\,)$ and $\xtwo = M^2 / (\xone  s)$, 
the latter yielding the value at which the parameter ${\hat q}$ should be evaluated~\cite{Baier:1996sk,Arleo:2012rs}. 
It is modelled by $\hat{q}(\xtwo) = \hat{q}_{_0} \left[{10^{-2}}/{\min(\xzero, \xtwo)} \right]^{_{0.3}}\!$~\cite{Arleo:2012rs}, 
with $\xzero = 1/(2 m_\tn{p} L_\tn{A})$, $\hat{q}_{_0} = 0.07\ \tn{GeV}^2/\tn{fm}$~\cite{Arleo:2020hat,Arleo:2021bpv}, 
and $L_\tn{A} =  \frac32\,r_{_0}\, A^{1/3}$ with $r_{_0}=1.12$~fm (but for the proton $L_\tn{p}=1.5$~fm~\cite{Arleo:2012rs}). 
We set $\xi=\frac12$ and use the procedure of Refs.~\cite{Arleo:2020eia,Arleo:2020hat,Arleo:2021bpv} to find a relative uncertainty on $R_\pA$ 
smaller than 10\%, see Fig.~\ref{fig:RpO}. 
Note that the strong coupling in \eq{dIR} should be evaluated at a scale 
$\sim \sqrt{\qhat  L_{\tn{A}}}$~\cite{Arleo:2012rs}  
which is semi-hard 
($\qhat  L_{\tn{A}} \sim 1 \, \tn{GeV}^2$, 
even at the smallest values of $\xtwo$ probed in the present study), 
and we have thus set $\alpha_s = 0.5$~\cite{Dokshitzer:1995qm,Courtoy:2013qca,Deur:2016tte} 
in our calculation.

%%%%%%%%%%%     RATIO R_{\nu}   %%%%%%%%%%%%%%%%%%

\begin{figure}[t]
  \includegraphics[scale=.83]{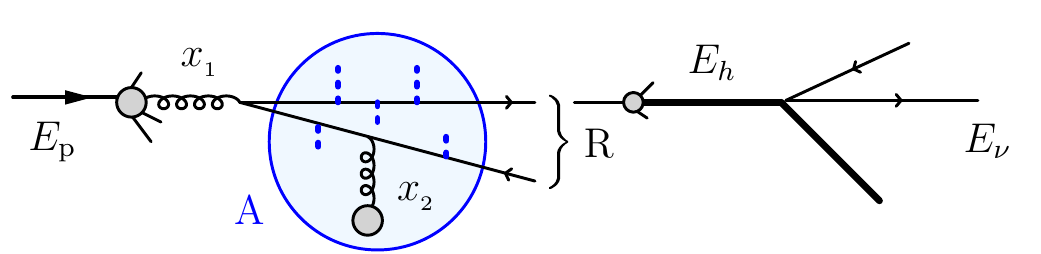}
  \caption{\label{fig:p-Air} 
  Particle interactions that result in a prompt neutrino on the Earth's surface, 
  viewed in the nucleus rest frame. 
  At LO, the $c{\bar c}$ pair 
  in colour representation R is produced by gluons  in the CR proton and in a target nucleon. 
  Fragmentation $c \to h$ occurs after the pair has undergone nuclear broadening
  and is typically followed by the hadron's semileptonic decay.
  }
\end{figure}

To make the connection between hadron nuclear suppression and 
the physics of atmospheric neutrino fluxes, 
note that $y=5$ in Fig.~\ref{fig:RpO} corresponds to a 
hadron energy in the oxygen rest frame $E_h \sim 10^6$~GeV, 
and we expect the neutrinos initiated by the decays of such hadrons 
(with energies $E_{\nu} \sim 0.6 \, E_h\,$\footnote{%
  In the case of the prompt neutrino flux, $E_\nu / E_D \sim 0.6$ can be inferred 
  from the decay $Z$-moment $Z_{D\nu}$ in Ref.~\cite{Lipari:1993hd}.
}) to be similarly diminished. 
However, the actual suppression of the {\it inclusive} neutrino flux 
(whose cascade production process is depicted in Fig.~\ref{fig:p-Air} in the case of prompt neutrinos), 
cannot be gleaned directly from Fig.~\ref{fig:RpO}. 
This is because the hadron flux samples the proton-air cross section over a wide range of proton energies 
(extending beyond the reach of the LHC), as recalled in what follows.

%%%%%%%%%%%%%%%%%%%%%%%%%%%%%%%%%%%%
%%%%%%%%%%%%%%%%%%%%%%%%%%%%%%%%%%%%%

The cascade equations needed to obtain the atmospheric neutrino flux $\Phi_\nu$ 
`deep in the atmosphere' can be solved approximately with 
the method of $Z$-moments~\cite{Lipari:1993hd,Gondolo:1995fq}. 
The gist of this approach is to factorize the initial proton interaction $\tn{p} + \tn{A}\to h + \tn{X}$ 
from the subsequent semileptonic hadron decay $h \to \nu\,$ as in Fig.~\ref{fig:p-Air}, 
leading to the neutrino flux approximated by 
(at an angle $\theta \lesssim 60^\circ$ from the zenith)~\cite{Lipari:1993hd,Gondolo:1995fq}
\be
\Phi_\nu(E_\nu) =  
  \frac{\Phi_\tn{p}(E_\nu)}{1-Z_\tn{pp}} \, \sum_{h} \, \frac{Z_{\tn{p} h} \, Z_{h\nu}}{1+B_h E_\nu  \cos \theta  / \varepsilon_h } \, .
\label{eq:phi-nu} 
\ee
Here $\Phi_\tn{p}$ is the initial CR flux, $\varepsilon_h$ is a critical energy 
(separating the low and high energy regimes for the flux of hadron $h$), 
and $Z_\tn{pp}$, $Z_{\tn{p} h}$, $Z_{h \nu}$ denote respectively the proton regeneration, 
hadron generation, and hadron decay $Z$-moments. 
The coefficient $B_h$ only depends on $Z_{h \nu}$ and on the nucleon and hadron attenuation lengths. 
Note that within the $Z$-moment approach, all $Z$-moments should 
have a smooth energy dependence and be evaluated at $E= E_\nu$ in Eq.~\eq{eq:phi-nu}.

Let us first focus on the prompt neutrino flux,  
$h=\{ D^\pm, D^0, D_s, \Lambda_c \}$ in Eq.~\eq{eq:phi-nu}. 
We will assume that $Z_{\tn{p} h} = f_h Z_{\tn{p} c}$ (see \eg\ ~\cite{Enberg:2008te,Goncalves:2017lvq}), 
where $f_h$ is the fraction from hadron $h$ to charm quark fragmentation~\cite{Yao:2006px}, 
and $Z_{\tn{p} c}$ is the {\it charm} generation $Z$-moment, which is itself proportional to 
an overlap integral $\Omega_\tn{pA}$ of the CR flux with the $\tn{p} + \tn{A} \to c + \tn{X}$ cross section~\cite{Gondolo:1995fq}, 
namely, 
\be
\Omega_\tn{pA}(E) \equiv
\int_0^1 \!  \frac{\dd \xF}{\xF} \, 
\Phi_\tn{p}\bigg(\frac{E}{\xF} \bigg) \, 
\frac{\dd \sigma_\tn{pA}^{c}}{\dd \xF} \bigg( \xF;\frac{E}{\xF}\bigg)  \, .
\label{eq:omega-def} 
\ee
This integral is the only quantity being sensitive to FCEL 
on which the neutrino flux~\eq{eq:phi-nu} (proportionally) depends.

Since the integral in Eq.~\eq{eq:omega-def} samples the charm cross section at different incoming proton energies 
rescaled by a factor $\xF$ \wrt~the energy $E$, and the FCEL effect rescales $\xF$ by a factor $z$, 
we easily find that $\Omega_\tn{pA}(E)$ is related to $\Omega_\tn{pp}(E)$ by a rescaling of the argument, 
\be
\Omega_\tn{pA}(E) = A \, \int_0^1 \! \dd z \, {\cal F}(z) \, \Omega_\tn{pp} \big( E/z \big) \, .
\label{eq:omegaAp}
\ee
Mind that in \eq{eq:omegaAp}, the quenching weight ${\cal F}(z)$ depends implicitly on 
the transport coefficient ${\hat q}(\xtwo)$ evaluated at $\xtwo = M^2 / (\xone s) \sim m_\perp^2 / (m_\tn{p} E)$, 
and thus fully determined by the neutrino energy $E$.

Appealing to Eqs.~\eq{eq:phi-nu}--\eq{eq:omegaAp}, we introduce the ratio $R_\nu$ of the 
neutrino flux calculated with FCEL to the version that assumes no nuclear effects, namely, 
\be 
R_{\nu}(E) \ \equiv \ 
\frac{1}{A} \, \frac{\Omega_\tn{pA}(E)}{\Omega_\tn{pp}(E)} 
= \int_0^1 \! \dd z \, {\cal F}(z) \, \frac{\Omega_\tn{pp} \big( E/z \big)} {\Omega_\tn{pp}(E)}  \, .
\label{eq:R_nu} 
\ee
(In the absence of nuclear effects, $\Omega_\tn{pA}(E) = A \, \Omega_\tn{pp}(E)$ and $R_{\nu} = 1$.) 
By focussing on $R_\nu$, we do not need 
to specify the other $Z$-moments and attenuation lengths that enter the expression \eq{eq:phi-nu} of the absolute 
neutrino flux.~A further 
virtue of nuclear ratios (like $R_{\rm pO}$) is that many systematic uncertainties 
are expected to cancel. 
We note that $\Omega_\tn{pp}(E)$ is a rapidly decreasing function of $E$. 
Indeed, in Eq.~\eq{eq:omega-def} moderate values of $\xF \simeq \xone /2$ 
dominate the integration and $\Phi_\tn{p}(E/\xF)$ thus decreases rapidly with increasing $E$, 
without being compensated by the much smoother increase of the pp cross section with $E\,$. 
The ratio $R_\nu$, being the average of ${\Omega_\tn{pp}(E/z)}/{\Omega_\tn{pp}(E)} < 1$ 
with the probability density ${\cal F}(z)$, will thus turn out to be a {\it suppression} factor. 

\begin{figure}[t]
  \includegraphics[scale=\figscale]{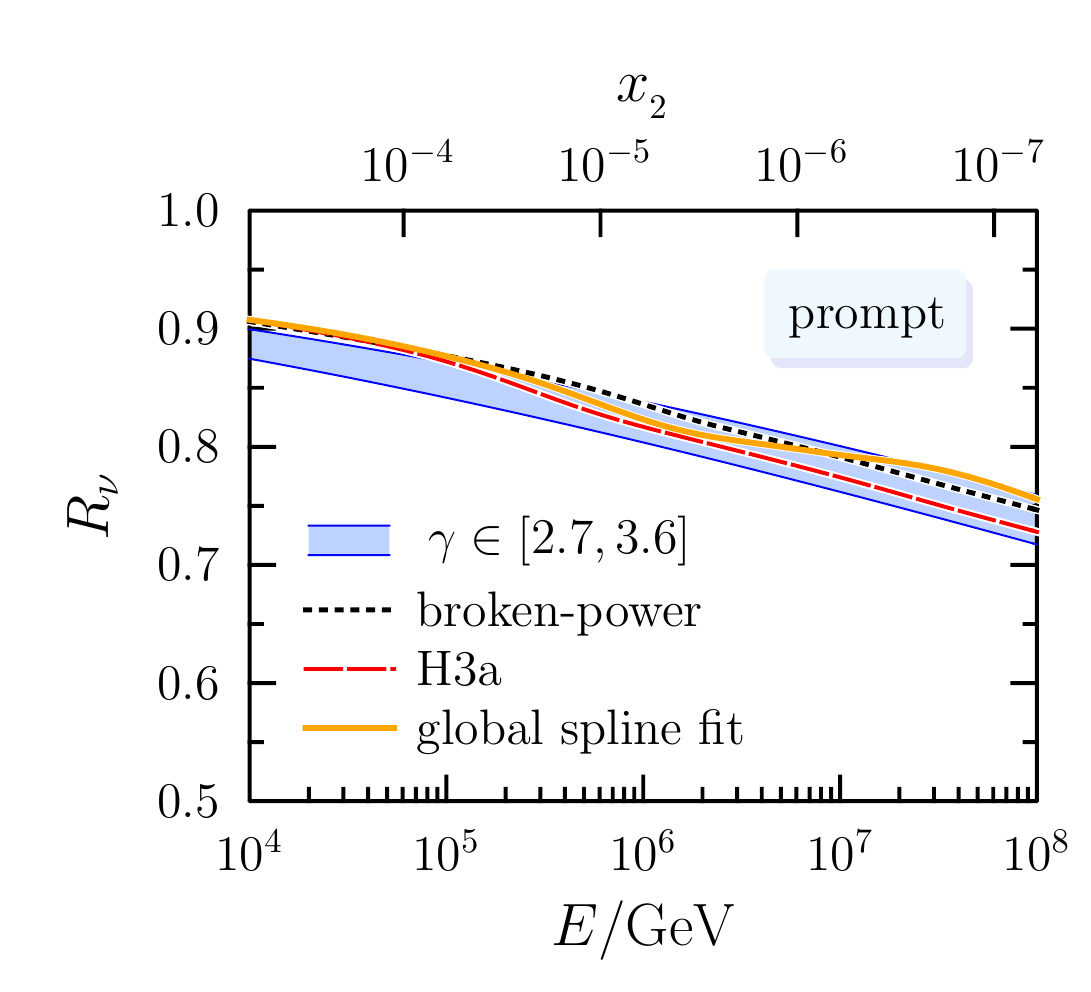}
  \vskip -3mm
  \caption{\label{fig:Rnu}
  The shaded region depicts Eq.~\eq{eq:scaling} when assuming 
  $\xF$-scaling of the pp cross section and a pure power law for the CR flux.
  The other curves are obtained from \eq{eq:R_nu} using various more realistic CR fluxes, 
  broken-power~\cite{Gondolo:1995fq}, H3a~\cite{Gaisser:2013bla} and GSF~\cite{Dembinski:2017zsh}, 
  as well as a more realistic pp cross section. 
  The corresponding values of $\xtwo$ are given on the top of the panel. 
  }
\end{figure}

It is instructive to first examine Eq.~\eq{eq:R_nu} assuming the ideal case where: 
(i) the CR flux is a pure power law, \ie\ 
$\Phi_\tn{p}(E) \propto E^{-\gamma}$  ($\gamma$ being commonly named the spectral index);
(ii) $\dd \sigma_{\tn{pp}}^c/\dd \xF$ does not depend on the energy and scales only with $\xF$.~Then 
both ${\Omega_\tn{pp}(E)}$ and ${\Omega_\tn{pp}(E/z)}$ defined by~\eq{eq:omega-def} are proportional 
to the Mellin transform 
$\int_{\tn{\tiny 0}}^{\tn{\tiny 1}} \dd \xF \, \xF^{\gamma-1} \, {\dd \sigma_\tn{pp}^c}/{\dd \xF}$ 
of the pp cross section \wrt~$\xF$, which thus cancels in~\eq{eq:R_nu}. 
Within the above scaling assumptions, 
the whole cascade process is uniquely determined by successive energy rescalings, 
and the adjustment due to FCEL quite naturally factorises. 
The resulting FCEL suppression factor of the neutrino flux is given by the $\gamma^\tn{th}$ moment of ${\cal F}(z)\,$, 
\be
R_\nu(E) = \int_{0}^1 \dd z \, z^{\gamma} \, {\cal F}(z) \, .
\label{eq:scaling}
\ee
As a consequence, $R_\nu$ depends on the charm production cross section only through the 
available colour channels in ${\cal F}$ and their probabilities, 
and on the neutrino energy only through ${\hat q}(\xtwo)\,$. 
The result \eq{eq:scaling} is represented in Fig.~\ref{fig:Rnu} between the values $\gamma = 2.7$ and $\gamma = 3.6$ 
bounding the spectral index in the considered energy range~\cite{Fedynitch:2018cbl}, 
corresponding respectively to the upper and lower boundaries of the filled band. 
Note that since $\hat{q}(\xtwo) \propto \xtwo^{-0.3}$ depends relatively smoothly on $\xtwo$ and enters the 
FCEL spectrum~\eq{dIR} through a logarithm, the ratio $R_\nu$ is a smooth function of energy, 
varying by only $\sim 25\%$ over four orders of magnitude. 
In Fig.~\ref{fig:Rnu} 
we used a typical value $K_\perp = 2$~GeV for charm production~\cite{Goncalves:2017lvq}. 
Varying $K_\perp$ in the range $[1.5,2.5]$~GeV gives a relative uncertainty on $R_\nu$ increasing from 2\% to 5\% for $E_\nu = 10^4...10^8$~GeV, 
which dominates the uncertainty associated to the set of parameters $\{ \hat{q}_{_0}, \xi, m, K_\perp \}$. 
%%%%%%%%%%%%%%%%%%%%%%%%%
%%%%%%%%%%%%%%%%%%%%%%%%%

In reality, the CR flux is not a pure power law 
(in other words,  $\gamma = \gamma(E_\tn{p})$~\cite{Fedynitch:2018cbl}) 
and $\dd \sigma_\tn{pp}/\dd \xF$ depends on $E_\tn{p}$ as $\dd \sigma_\tn{pp}^c / \dd \xF \simeq a(E_\tn{p}) f(\xF)$, 
with $a(E_\tn{p})$ a smoothly increasing function of $E_\tn{p}$. 
When implementing \eq{eq:pA} into \eq{eq:omega-def}, the factor $a(E_\tn{p})$ can be absorbed by 
slightly shifting down the spectral index $\gamma(E_\tn{p})$ of $\Phi_\tn{p}$.~Thus, 
the calculation of \eq{eq:R_nu} using 
a more realistic pp cross section and various existing models of the CR flux 
should be well encompassed by the ideal situation considered above 
(where $\gamma$ varies in its customary range), 
up to slight deviations in the energy domains where the spectral index is smallest. 
This is confirmed in Fig.~\ref{fig:Rnu} by the results corresponding to various choices 
of the CR flux with the parametrization for $\dd \sigma_\tn{pp}^c / \dd \xF$
from Ref.~\cite{Martin:2003us}.
Although the precise form of $f(\xF)$ in the latter parametrization 
is quite irrelevant~\footnote{%
  In the parametrization of Ref.~\cite{Martin:2003us}, 
  $\dd \sigma_\tn{pp}^c / \dd \xF = a(E_\tn{p}) f(\xF, E_\tn{p})$, 
  the function $f$ has a slight energy dependence, which 
  has negligible effects in our study and can thus be ignored, $f(\xF, E_\tn{p}) \simeq f(\xF)$.
}, 
its Mellin transform tells us about the typical $\xF$ in the cascade. We find $\langle \xF \rangle \sim 0.1$ 
 and hence $\langle \xone \rangle \sim 0.2$, which is 
fully consistent with the assumption that $gg \to Q\bar{Q}$ dominates over $q \bar{q} \to Q\bar{Q}$
(in particular since $\xtwo < 10^{-3}$ in the energy range of Fig.~\ref{fig:Rnu}).

Our calculation of $R_\nu$ for the prompt flux can be easily adapted for the conventional flux, 
by summing over light hadrons in Eq.~\eq{eq:phi-nu}, 
$h=\{\pi^\pm, K^\pm, K^0_L \}$, 
and considering the LO partonic process $qg \to qg$ to be dominant. 
Hence colour states and probabilities are changed, 
modifying the FCEL spectrum \eq{dIR} (with $C_1 = C_F\,$, 
$C_2 = N_c$, $\R = \{\bm 3, \bm{\bar 6}, \bm{15} \}$) 
and the quenching weight ${\cal F}(z)\,$. 
In the particular case where the quark-gluon pair is a colour triplet, $\R = \bm 3$, 
the spectrum~\eq{dIR} is negative and the rescaling factor $z>1$~(see Ref.~\cite{Arleo:2020hat}).
%
%%%%%%%%%%%%%%%%%%%%%%%%%
\begin{figure}[t]
 \includegraphics[scale=\figscale]{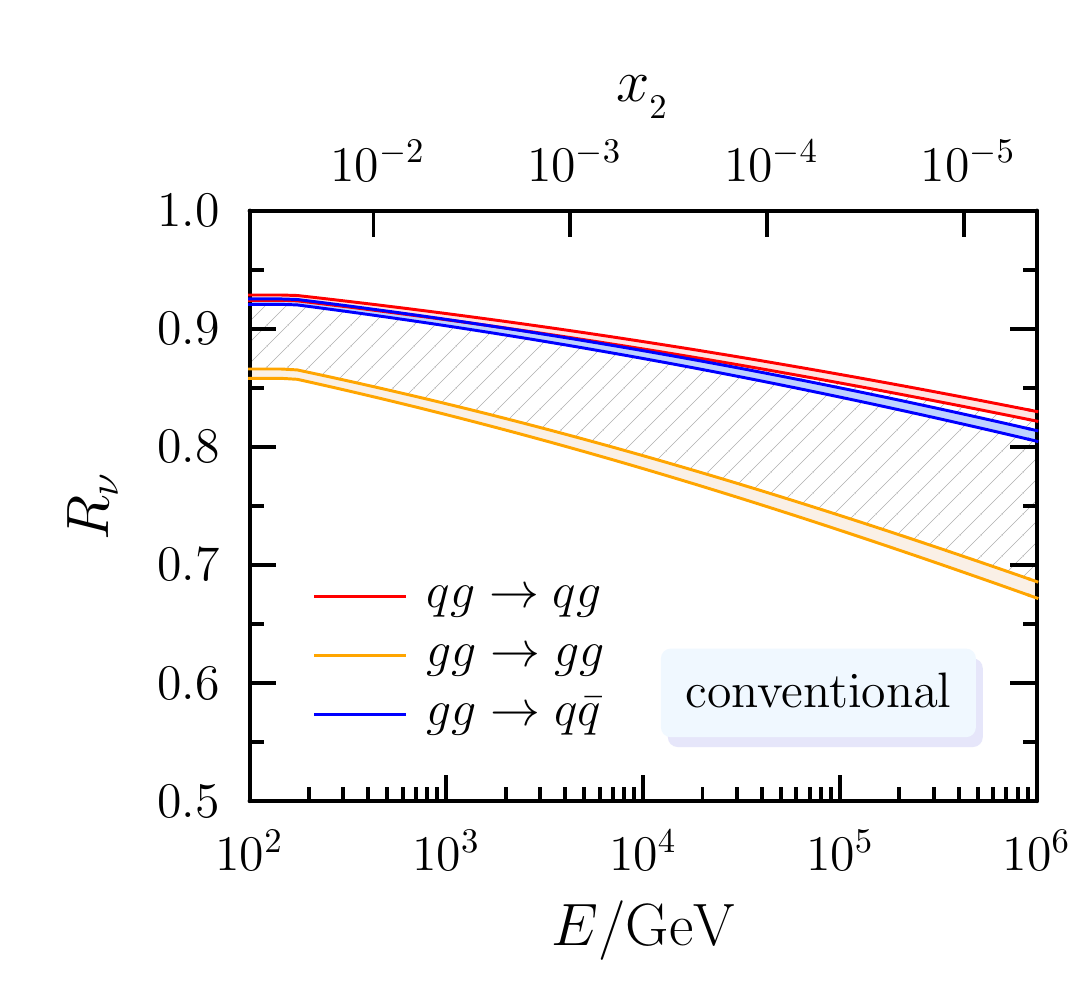}
  \vskip -3mm
  \caption{\label{fig:Rconv} 
  Like Fig.~\ref{fig:Rnu}, but for the conventional flux in a lower neutrino
  energy range. 
  For each partonic channel, the narrow band arises from 
  varying $\gamma$ between $\gamma = 2.7$ and $\gamma = 2.9$.
  The channel $qg \to qg$ is expected to be prevalent 
  in the region of momentum fractions where the PDFs are evaluated.
  }
\end{figure}
%%%%%%%%%%%%%%%%%%%%%%%%%
The result for $R_\nu$ is shown as the uppermost band in Fig.~\ref{fig:Rconv},
which is narrow because the spectral index is taken as $2.7 < \gamma < 2.9$ in 
the reckoned energy range $10^2...10^6$~GeV~\cite{Fedynitch:2018cbl}. 
Here we used $f(\xF) = (1-\xF)^{3}/\xF$ in the pp cross section parametrization for light hadron production, based on forward $\pi^0$ production  
at the LHC~\cite{Adriani:2015iwv}. 
Calculations corresponding to more realistic CR fluxes and pp cross section parametrizations 
all fall within this narrow band. 
Compared to the prompt calculation, we find a larger typical $\langle \xF \rangle \sim 0.2$, 
consistent with the assumption that $qg \to qg$ takes precedence 
(over $gg \to q \bar{q}$ and $gg \to gg$) for the conventional neutrino flux, 
the valence quark PDF being larger 
than the gluon PDF at $\langle \xone \rangle \sim 0.4\,$.  
However, we also show in Fig.~\ref{fig:Rconv} 
the results for $R_\nu$ assuming the processes $gg \to q \bar{q}$ and $gg \to gg$.
Due to larger Casimir charges accessible for those processes, 
the resulting ratio $R_\nu$ is smaller. 
Accounting for the contribution of those channels in a full perturbative QCD 
calculation would thus tend to reduce $R_\nu\,$.
In Fig.~\ref{fig:Rconv}, the prediction assuming $qg \to qg$ should therefore 
be viewed as an upper estimate of $R_\nu$ 
and the hatched area as a very conservative uncertainty band. 
Here we also took $K_\perp = 2$~GeV, whose variation 
by $\pm 0.5$~GeV only leads to a relative uncertainty smaller than $10$\% 
for each partonic process.

%%%%%    DISCUSSION     %%%%%%%

In summary, 
motivated by the nuclear suppression in pO collisions at the LHC expected from FCEL 
(see Fig.~\ref{fig:RpO}), 
we have quantified the influence of this effect on atmospheric neutrinos through 
the ratio $R_\nu$, Eq.~\eq{eq:R_nu}.
The flux is depleted, and the reduction proves more pronounced with increasing energy 
(see Figs.~\ref{fig:Rnu} and \ref{fig:Rconv}), independently of the model used for the 
initial CR flux and hadron production cross section in proton-air collisions. 
Several QCD studies have already focused on nuclear effects for prompt neutrinos, 
from the perspective of either nuclear parton distribution functions 
(nPDFs) or gluon saturation at small $\xtwo$~\cite{Martin:2003us,Henley:2005ms,Enberg:2008te,Garzelli:2016xmx,Bhattacharya:2015jpa,Garzelli:2015psa,Bhattacharya:2016jce,Benzke:2017yjn}. 
Compared to nPDF effects, which suppress the flux by 
$\sim 10...30\,\%$ (depending on $E_\nu$ and on the nuclear PDF set), 
the FCEL effect is at least 
as important (see Fig.~\ref{fig:Rnu}) and brings the prompt neutrino 
flux further below the experimental upper bound
determined by the IceCube collaboration~\cite{Aartsen:2016xlq}.
There are also calculations of the conventional flux~\cite{Honda:2006qj,Fedynitch:2018cbl}, 
aiming to reduce the absolute flux uncertainty to the sub-10\% level, 
while the FCEL effect is expected to be at least $\sim 10\%$ in that case (see Fig.~\ref{fig:Rconv}). 

The present prospects on theoretical and experimental accuracies 
make it timely to involve FCEL in predictions of both kinds of atmospheric neutrino sources. 
Moreover, FCEL should not only affect inclusive fluxes, 
but also the structure of air showers, and it would seem pertinent to incorporate FCEL 
in air shower simulations and Monte Carlo event generators 
(\eg\ CORSIKA \cite{Engel:2018akg}, EPOS~\cite{Pierog:2013ria}, QGSJET-III~\cite{Ostapchenko:2019few}, Sibyll \cite{Fedynitch:2018cbl,Riehn:2019jet}).
This would also remove some unavoidable theoretical uncertainty of the $Z$-moment method adopted in this Letter, 
since it is inclusive in kinematic variables such as $K_\perp$ and $\xi\,$. 
Using a full air shower simulation would remove such ambiguities, 
thus improving the accuracy of atmospheric neutrino fluxes accounting for the FCEL effect. 

This work is funded 
by the ``Agence Nationale de la Recherche'' under grant ANR-COLDLOSS (ANR-18-CE31-0024-02) 
and
by the U.S. Department of Energy (DOE) under grant No.~DE-FG02-00ER41132. 


%\bibliography{mybib}
%\bibliographystyle{apsrev4-2}

\end{document}